\documentclass[preprint,12pt,authoryear]{elsarticle}
\usepackage{subcaption}
\usepackage{booktabs}
\usepackage[figuresright]{rotating}
\usepackage{amsmath,amssymb,amsfonts}
\usepackage{algorithmic}
\usepackage{graphicx}
\usepackage{multirow}

\usepackage{amssymb}

\begin{document}

\begin{frontmatter}



\title{Chronic Disease Diagnoses Using Behavioral Data}

\author[label1]{Di Wang$^*$}
\author[label1]{Yidan Hu$^*$}
\affiliation[label1]{organization={Nanyang Technological University},
            addressline={Singapore}, 
            }
\author[label2]{Eng Sing Lee}

\affiliation[label2]{organization={National Healthcare Group Polyclinics},
            addressline={Singapore}, 
            }
\author[label3]{Hui Hwang Teong}
\author[label3]{Ray Tian Rui Lai}
\affiliation[label3]{organization={Tan Tock Seng Hospital},
            addressline={Singapore}, 
            }

\author[label4]{Wai Han Hoi}
\affiliation[label4]{organization={Woodlands Health},
            addressline={Singapore}
            }
\author[label1]{Chunyan Miao}

\begin{abstract}
Early detection of chronic diseases is beneficial to healthcare by providing a golden opportunity for timely interventions. Although numerous prior studies have successfully used machine learning (ML) models for disease diagnoses, they highly rely on medical data, which are scarce for most patients in the early stage of the chronic diseases. In this paper, we aim to diagnose hyperglycemia (diabetes), hyperlipidemia, and hypertension (collectively known as 3H) using own collected behavioral data, thus, enable the early detection of 3H without using medical data collected in clinical settings. Specifically, we collected daily behavioral data from 629 participants over a 3-month study period, and trained various ML models after data preprocessing. Experimental results show that only using the participants' uploaded behavioral data, we can achieve accurate 3H diagnoses: 80.2\%, 71.3\%, and 81.2\% for diabetes, hyperlipidemia, and hypertension, respectively. Furthermore, we conduct Shapley analysis on the trained models to identify the most influential features for each type of diseases. The identified influential features are consistent with those reported in the literature.



\end{abstract}







\end{frontmatter}

\def\thefootnote{*}\footnotetext{These authors contributed equally to this work}\def\thefootnote{\arabic{footnote}}

\section{Introduction}\label{sec_intro}

Unhealthy lifestyles lead to high incidences of chronic diseases, which is happening pervasively around the globe. For instance, in Singapore, more than 25\% of people over the age of 40 suffer from at least one type of chronic disease (\cite{hpbTipsPrevent}). Chronic diseases mainly damage essential organs such as the brain, heart, and kidney, and are the leading causes of disability and death. Among many chronic diseases, hyperglycemia (diabetes), hyperlipidemia, and hypertension are probably the three most common ones, which are often referred together as 3H. Globally speaking, the predicted number of adult patients with diabetes will rise from 415 million in 2015 to 642 million by 2040 \cite{ogurtsova2017idf}. The prevalence of hyperlipidemia and hypertension among adults is both high that about 39\% suffered from hyperlipidemia back in 2008 \cite{al2021prevalence} and 31.1\% suffered from hypertension back in 2010 \cite{mills2016global}. Early detection of chronic diseases, especially the 3H, is vitally important to improve a vast amount of persons' quality of life and alleviate the economic burden due to high medical costs.

Many prior studies have been conducted to diagnose 3H diseases using medical data. For instance, Machine Learning (ML) classification algorithms were used to predict diabetes \cite{sisodia2018prediction, choubey2017classification} on the Pima Indians Diabetes Dataset (PIDD), which comprises blood glucose and blood pressure data collected from hospitals. A predictive model was trained to diagnose hypertension \cite{ye2018prediction} using the Maine Health Information Exchange Network dataset, which comprises demographic and medical data. Electronic medical records were leveraged to predict hyperlipidemia using ML models \cite{liao2022dhdip}.

However, medical data are expensive, thus, often not comprehensively available for majority people. On one hand, medical data are mostly collected in clinical settings by professionally trained persons (e.g., doctors, nurses, occupational healthcare providers, etc.). Such data collection often incurs relatively high monetary and time costs. On the other hand, there is a prevalent stereotype, especially in Asian countries, that people have a negative view towards visiting hospitals. Many will only do so when their health status has deteriorated badly enough that visiting hospitals is unavoidable. As a result of these leading factors, during the early stage of most chronic diseases, especially the 3H, patients are often unaware that they have a great risk of developing or are already suffering from these diseases due to the subtle symptoms.

Because the early detection of chronic diseases is critical, however, challenging in clinical settings, healthcare professionals and ML researchers are working together towards detecting the early symptoms of chronic diseases using behavioral data. Such choice is well grounded because other than genotype and external factors, most chronic diseases are the long-term outcome of ones' lifestyle. The authors of this paper comprise both ML researchers and medical doctors. Together, we developed a mobile app to collect over 600 users' daily behavioral data for a 3-month study period (see Section~\ref{sec_data}) and subsequently trained a set of ML algorithms (see Section~\ref{sec_ML}) to answer the following research question:

\textbf{\emph{Whether and to what extent the collected daily behavioral data can be used to accurately assess one's 3H status?}}

Before we attempt to answer such research question by applying ML algorithms onto the behavioral dataset, we have to preprocess the collected data mainly due to the following two reasons. Firstly, akin to most real-world datasets, especially in the field of healthcare, there are always missing values. In this paper, we apply both Mean Imputation (MI) and $k$-Nearest Neighbor Imputation (KNNI) methods to handle the missing values (see Section~\ref{sec_process_data_A}). Secondly, because the behavioral data were uploaded by users without anyone's supervision, we need to carefully examine such self-reported data and remove any detected false data entries (see Section~\ref{sec_process_data_A}). Furthermore, because the study spanned across three months that one's lifestyle may already deviate, we purposely extract features to represent such changes in behavior (see Section~\ref{sec_process_data_A}). As the Shapley explainability assessments suggest (see Section~\ref{sec_exp_shap}), such extracted behavioral changes do significantly contribute to the overall disease diagnoses. Experimental results suggest that it is quite accurate to diagnose 3H diseases using only the collected daily behavioral data. Specifically, the accuracy of diagnosing diabetes and hypertension is both over 0.8 and that of diagnosing hyperlipidemia is over 0.7 (see Section~\ref{exp_result}). Furthermore, it is extremely encouraging to learn that using one's comprehensive daily behavioral data is more accurate in the diagnosis of hypertension than only relying on the traditional way of taking a series of blood pressure measurements (see Fig.~\ref{figure:hyper_acc}). In summary, we adequately answer the research question by showing that we can accurately diagnose 3H diseases by using the collected comprehensive daily behavioral data only.

The contributions of this research work are as follows.
\begin{itemize}
\item We introduce how we collected daily behavioral data and preprocessed them for subsequent disease diagnoses.
\item We conduct extensive experiments on 3H disease diagnoses. Experimental results show that we can rely on the collected behavioral data for accurate diagnoses of 3H. Notably, in terms of hypertension, our approach outperforms the traditional way of taking a series of blood pressure measurements.
\item We investigate feature importance with the aid of Shapley values and find the identified influential features highly consistent with those reported in the literature.
\end{itemize}



\section{Related Work} \label{sec_RW}
In this section, we review two lines of prior studies, i.e., ML models used for 3H diagnoses and the related datasets.

\subsection{Machine Learning Models Used for 3H Diagnoses}
Diabetes, hyperlipidemia, and hypertension, collectively known as 3H, are typical chronic diseases that may greatly affect one's quality of life and even reduce life expectancy \cite{zavoreo2012triple}. Early detection and timely intervention are essential to prevent 3H from deterioration or even to reverse the 3H status. Recently, many studies have proposed to predict these chronic diseases using various ML models.

To diagnose diabetes, Kavakiotis et al. \cite{kavakiotis2017machine} studied various ML algorithms in diabetes prediction, diabetic complications, genetic background and environment, and healthcare and management. 
Sisodia and Sisodia \cite{sisodia2018prediction} diagnosed diabetes with Decision Tree (DT), Support Vector Machine (SVM), and Naive Bayes (NB) on the Pima Indians Diabetes Database (PIDD) and evaluated the performance with precision, accuracy, F-measure, and recall. NB achieved the highest accuracy in their study.
Zou et al. \cite{zou2018predicting} conducted feature reduction using Principal Component Analysis (PCA) and minimum Redundancy Maximum Relevance (mRMR) and utilized Random Forest (RF), J48, and Artificial Neural Network (ANN) for binary diabetes prediction. In their study, RF outperformed the others. Tigga and Garg \cite{tigga2020prediction} collected participants' health, lifestyle, and family background data using online and offline questionnaires and predicted diabetes with logistic regression, K-Nearest Neighbour (KNN), SVM, NB, DT and RF. Among these, RF was again found to be the most accurate.

To diagnose hyperlipidemia, Berina et al. \cite{berina2021using} leveraged ANN to assist doctors in diagnosing familial hyperlipidemia. Zhang et al. \cite{zhang2019automatic} proposed a diagnostic system utilizing data extension and data correction, which can diagnose hyperlipidemia with human physiological parameters comprising blood glucose detection, glycosylated hemoglobin test, routine blood examination, routine urine test, and biochemical detection. Liu et al. \cite{liu2020deep} proposed an auxiliary diagnosis support algorithm to diagnose hyperlipidemia with text-based medical data based on a deep learning algorithm with attention.

To diagnose hypertension, Zhu et al. \cite{zhu2021prediction} leveraged the Long Short-Term Memory (LSTM) model and Bayesian fitting method using long-term indoor environmental data. Kim et al. \cite{kim2022classification} used a deep
neural network to study the effect of energy intake adjustment in hypertension. Leha et al. \cite{leha2019machine} used RF of various types of trees namely classification, regression, lasso penalized logistic regression, and boosted classification trees, and SVM to predict Pulmonary Artery Pressure (PAP) with echocardiographic estimations of PAP obtained within 24 hours. Their results show that the RF of regression trees achieves the best performance in terms of the AUC metric. 

These prior studies focus on one type of disease. However, 3H is a well known comorbidity problem, which means there is a high prevalence of having more than one diseases among the three \cite{petrie2018diabetes}. In this paper, we diagnose all three of them and investigate the feature importance regarding different diseases.

\subsection{3H Disease Diagnosis Datasets}
Apart from ML algorithms, the input data used are also critical to 3H diagnoses. For the diagnosis of diabetes, the most widely used dataset is PIDD 
, which comprises 768 instances of female participants over the age of 21. There are eight features in PIDD, namely age, BMI, glucose, insulin level, Blood Pressure (BP), number of pregnancies, skin thickness, and outcome label, out of which most are medical data. 
The Singapore Epidemiology of Eye Disease (SEED) dataset has also been used for diabetes diagnosis \cite{tan2018ethnic}. SEED comprises 10,033 patients' records with profile data such as income, education, and retinopathy attributes as enrichment.

For the diagnosis of hypertension, both the Maine Health Information Exchange Network dataset \cite{ye2018prediction} and the Henry Ford Health Systems database \cite{sakr2018using} have been widely used \cite{hao2015development, elshawi2019interpretability}. They both comprise participants' profile data (e.g., age and gender) and comprehensive medical data. Moreover, the Beijing Chinese Han population dataset, which used the epidemiological investigation questionnaire to collect 1,200 patients' records and included environmental and genetic factors, has also been used for hypertension diagnosis \cite{pei2018risk}.

So far, to the best of our knowledge, there is no publicly available dataset designated for the diagnosis of hyperlipidemia (note that \cite{liao2022dhdip} reviewed in Section~\ref{sec_intro} is not meant for hyperlipidemia diagnosis only). One possible reason is because such disease can be straightforwardly diagnosed using a conventional blood test. There seems to be a lack of promising results on the diagnosis of hyperlipidemia using behavioral data up until now. Nevertheless, the Jackson Heart Study (JHS) dataset \cite{wyatt2008prevalence} is a comprehensive collection of 3,340 participants’ profile data, lifestyle data, medical measurement data, and other sociocultural attributes, which may be used for the diagnoses of all 3H diseases \cite{sims2011socioeconomic, thallapureddy2003lipid}.

Models trained on the reviewed datasets usually perform well, mainly because these datasets are rich in various medical measurements and have (relatively) few missing values. However, as discussed in Section~\ref{sec_intro}, medical data are most likely unavailable for the early detection of 3H among a large population. In this paper, we aim to diagnose 3H diseases using only the collected daily behavioral data instead of the comprehensive medical data collected in clinical settings.

\section{Daily Behavioral Dataset for 3H Diagnosis} \label{sec_data}
In this section, we introduce the details of our collected {Singaporean daily behavioral dataset for 3H Diagnosis (\textbf{S3D}).

As introduced in Section~\ref{sec_intro}, for the early detection of 3H diseases, we hypothesize to collect longitudinal behavioral data and use such data for accurate 3H diagnosis instead of using medical data collected in clinical settings. Towards this goal, we developed a mobile app for users to upload their daily behavioral data, including sleep hours, step counts, daily activities they perform (see Table~\ref{activity_summary}) with respective duration, etc. If the recruited users (participants of the study) has their own blood pressure monitor and/or glucometer at home, we encourage them to upload their blood pressure (BP) readings (ideally twice a day, eight hours apart) and/or blood glucose (BG) readings (ideally twice a day, once before meal and the other two hours after meal). Although BP and BG readings may not be conventionally considered as behavioral data, because these readings are self-tested at home (not in a clinical environment) by those participants who concern much on their health status, we deem these readings as uploaded daily behavioral data. To ease data collection, we enable the app to link with wearables (e.g., Fitbit or another mobile app) for automated entries of step counts, sleep duration, and activities, and to capture the BP or BG reading autonomously by taking a photo of the meter. Nonetheless, after linking the activity data and/or capturing the meter reading, the participant is allowed to make amendments because the linked and captured data may not always be accurate. Therefore, we still consider all the data uploaded through our app as self-reported by the participants, hence, there is a need to validate the authenticity of such data (see Section~\ref{sec_data_clean}).

Our study had obtained the relevant ethics approval. Altogether, we recruited 629 participants (mean age=57.96, std=13.23) from hospitals, clinics, and community centres from end 2021 to mid 2022. All participants were asked to use our app to upload their daily behavioral data for three consecutive months. Among these 629 participants, majority are female (402, 63.91\%). The spread of ethnic groups well represents the respective ratio of Singapore that in our study, majority participants are Chinese (493, 78.38\%), followed by Malay (67, 10.65\%), Indian (54, 8.59\%) and others (15, 2.38\%). The disease labels (see Table~\ref{tab_3H_ratio}) as well as the basic profile data (i.e., BMI, smoking and drinking status, see Table~\ref{tab:attributes}) are obtained upon participant recruitment. We will make S3D public once this paper is accepted for publication and the relevant approval is obtained.

\begin{table}[!t]

\caption{3H status among recruited participants}
\centering
\begin{tabular}{lrrr}
\hline
Disease        & Yes & No & Pre-disease\\ \hline
Diabetes       & 244 (38.79\%)    & 349 (55.49\%) & 36 (5.72\%)                                                                               \\
Hyperlipidemia         & 364 (57.87\%)     & 265 (42.13\%)& N.A.                                                                               \\
Hypertension         & 350  (55.64\%)   & 279   (44.36\%)  & N.A.                           \\ \bottomrule
\end{tabular}
\label{tab_3H_ratio}
\end{table}

\begin{sidewaystable}
\centering
\scriptsize
\caption{Attributes in S3D dataset}
\begin{tabular}{@{}lllcl@{}}
\toprule
Index & Attribute                 & Descriptions                                            & Range  & MV Numbers (ratio)              \\ \midrule
1     & Gender                    & female, male                                            & -      & 0 (0.00\%)              \\\hline
2     & Age                       & current age                                             & -      & 0 (0.00\%)              \\\hline
3     & Ethnic group           & Chinese, Malay, Indian, Eurasian, Philippines,   Others & -      & 0 (0.00\%)              \\\hline
4     & BMI                       & body mass index in kg/m2                                & {[}16.44, 517.82{]}  & 0 (0.00\%)\\\hline
5     & Smoking status            & non-smoker, ex-smoker, smoker                           & -      & 0 (0.00\%)              \\\hline
6     & Drinking status           & non-drinker, ex-drinker, drinker                        & -      & 0 (0.00\%)              \\\hline
7     &  BG             &   blood glucose in mmol/L                                          & {[}4.0, 29.0{]}  & 346 (68.79\%)    \\\hline
8     & $\text{BG}_\textit{BM}$             &    blood glucose before meal                                                     & {[}4.0, 13.27{]}    &360 (71.57\%)\\\hline
9     & $\text{BG}_\textit{AM}$         &    blood glucose two hours after meal                                                     & {[}4.71, 29.0{]}   &394 (78.33\%) \\\hline
10    & $\text{BG}_{C}$ &   gap value of blood glucose between $\text{BG}_\textit{BM}$ and $\text{BG}_\textit{AM}$                                                      & {[}0.0, 9.77{]} &408 (81.11\%)     \\\hline
11    & SBP                       & systolic blood pressure in mmHg                         & {[}95.6, 163.5{]}  &154 (30.62\%)     \\\hline 
12    & $\text{SBP}_{F}$    & participants' SBP during the former half of participation time                                                      & {[}96.0, 167.0{]}  &154 (30.62\%)  \\\hline
13    & $\text{SBP}_{L}$      &  participants' SBP during the latter half of participation time                                                      & {[}92.0, 175.5{]} &192 (38.17\%)   \\\hline
14    & $\text{SBP}_{C}$    &   gap value of blood pressure between $\text{SBP}_{F}$ and $\text{SBP}_{L}$                                                         & {[}0.0, 38.0{]}   &192 (38.17\%)    \\\hline
15    & DBP                       & diastolic blood pressure in mmHg                        & {[}46.0, 111.0{]} &154 (30.62\%)   \\\hline
16    & $\text{DBP}_{F}$   & participants' DBP during the former half of participation time                                                      & {[}46.0, 111.0{]}   &154 (30.62\%) \\\hline
17    & $\text{DBP}_{L}$     & participants' DBP during the latter half of participation time                                                       & {[}56.78, 99.67{]}  &192 (38.17\%)  \\\hline
18    & $\text{DBP}_{C}$   &  gap value of blood pressure between $\text{DBP}_{F}$ and $\text{DBP}_{L}$                                                       & {[}0.0, 25.0{]}   &192 (38.17\%)    \\\hline
19    & Step                &  mean value of all uploaded step numbers              & {[}1.0, 27437.52{]}  &33 (6.56\%) \\\hline
20    & $\text{Step}_{F}$    &   mean value of all uploaded step numbers during the former half of participation time                                                        & {[}1.0, 27827.38{]} &33 (6.56\%)\\\hline
21    & $\text{Step}_{L}$    &   mean value of all uploaded step numbers during the latter half of participation time                                                           & {[}1.67, 42608.70{]} &43 (8.55\%) \\\hline
22    & $\text{Step}_{C}$  &    gap value of step numbers between $\text{Step}_{F}$ and $\text{Step}_{L}$                                                       & {[}0.0, 36965.22{]}  &43 (8.55\%)  \\\hline
23    & Step Count               & count of times for uploading steps                          & {[}1.0, 117.0{]}     &33 (6.56\%)\\\hline
24    & $\text{Step}_{F}$ Count   & count of times for uploading steps during the former half of participation time                                                       & {[}1.0, 58.0{]}   &33 (6.56\%)   \\\hline
25    & $\text{Step}_{L}$ Count    &  count of times for uploading steps during the latter half of participation time                                                       & {[}1.0, 59.0{]}  &43 (8.55\%)     \\\hline 
26    & $\text{Step}_{C}$ Count &   gap value of step count between $\text{Step}_{F}$ and $\text{Step}_{L}$                                                        & {[}0.0, 1.0{]}   &43 (8.55\%)     \\\hline
27    & Sleep                     &  mean value of all uploaded sleep minutes            & {[}45.0, 604.02{]}   &76 (15.11\%) \\\hline
28    & $\text{Sleep}_{F}$       &   mean value of all uploaded sleep minutes during the former half of participation time   & {[}1.0, 115.0{]}   &76 (15.11\%)  \\\hline
29    & $\text{Sleep}_{L}$         &   mean value of all uploaded sleep minutes during the latter half of participation time   & {[}30.0, 731.5{]}   &110 (21.87\%)   \\\hline
30    & $\text{Sleep}_{C}$          &   gap value of sleep minutes between $\text{Sleep}_{F}$ and $\text{Sleep}_{L}$        & {[}35.83, 660.0{]}  &110 (21.87\%) \\\hline
31    & Sleep Count              & count of times for uploading sleeps minutes                        & {[}0.0, 378.0{]}  &76 (15.11\%)   \\\hline
32    & $\text{Sleep}_{F}$ Count      & count of times for uploading sleep minutes during the former half of participation time    & {[}1.0, 57.0{]}   &76 (15.11\%)   \\\hline
33    & $\text{Sleep}_{L}$ Count  & count of times for uploading sleep minutes during the latter half of participation time                                                          & {[}1.0, 58.0{]}     &110 (21.87\%) \\\hline
34    & $\text{Sleep}_{C}$ Count         & gap value of sleep count between $\text{Sleep}_{F}$ and $\text{Sleep}_{L}$                                                        & {[}0.0, 1.0{]}   &110 (21.87\%)    \\\hline
35    & Activity             & physical activity level (see (\ref{equ:fusion}))                            & {[}0.0, 245679.5{]}  &26 (5.17\%) \\\hline
36    & Diabetes                  & pre-diabetes (pre-DM), diabetes mellitus (DM), and no DM                                       & -     & 0 (0.00\%)               \\\hline
37    & Hyperlipidemia             & yes if diagnosed with hyperlipidemia, otherwise   no & -   & 0 (0.00\%)                 \\\hline
38    & Hypertension              & yes if diagnosed with hypertension, otherwise no    & -    & 0 (0.00\%)                \\ \bottomrule
\end{tabular}

\label{tab:attributes}
\end{sidewaystable}

\section{Using S3D Dataset for 3H Diagnosis}\label{sec_process_data}

In this section, we present the data processing steps and introduce the ML algorithms used for 3H diagnosis.

\subsection{Data Preprocessing} \label{sec_process_data_A}
 Data preprocessing is essential to standardize the dataset for model training. In this subsection, we introduce how we perform data cleaning, augmentation, and imputation.


\subsubsection{Data Cleaning} \label{sec_data_clean}
In this study, all the 629 participants enrolled for approximately three months. To ensure high-level data quality to train meaningful models, we first remove a participant’s data if he/she uploaded data via our mobile app for less than ten days during the 3-month study (113 such participants), and if his/her profile data have critical missing values (such as age, gender, and BMI, 6 such participants). After such removals, we have the daily behavioral data of 510 participants. Furthermore, although our system is able to auto-load/recognize certain data (such as step counts, sleep hours, measurement readings, etc.), the participants may still choose to self-report/alter these readings. Thus, there is a need to further check whether there are any suspicious self-reported data that may highly likely be intentionally repeatedly entered by a few participants on different days, because doing this may quickly help them to complete the daily task(s) given by our mobile app. Upon the examination of repeated data entries on different days, we find a small group of participants who entered identical sleep hours for more than ten days (i.e., std=0 for all their daily sleep hours). Because the report of sleep hours has a granularity of minutes (e.g., 7 hours and 20 mins), we deem having exactly the same sleep duration for more than ten days as highly unlikely, and thus remove the data of these 7 participants. In the end, we have a dataset comprising 503 participants’ daily behavioral data, i.e., 116 participants’ data (20.03\%) are removed.

\subsubsection{Feature Engineering}
As shown in Table ~\ref{tab:attributes}, we use a total number of 38 medical, profile, and activity attributes for 3H diagnoses. The last three attributes, i.e., Diabetes, Hyperlipidemia, Hypertension, are the labels of 3H, respectively. Apart from few attributes remaining invariant throughout the 3-month study, namely gender and ethnic group, the others evolved along with new data being uploaded by the participants, especially for medical measurement data (e.g., blood pressure) and activity data (e.g., step count). For those attributes having varying values, we utilize the mean value across the entire study period to represent the overall value. However, only using the mean value will overlook the dynamic changes during the 3-month study period, which might be critical for revealing the changes in participants' disease status. To effectively capture such dynamic changes while still maintaining simplicity, we conduct data augmentation on these attributes. Specifically, we divide the data uploaded by each participant into two subsets evenly by following their chronological order and compute two mean values, respectively. We further compute the difference between the two mean values as another attribute for the relevant change. We denote the measurement for the former half and the latter half using subscripts $F$ and $L$, respectively, and denote the change in between using subscript $C$. For instance, one's systolic blood pressure for the former half period of the study is denoted as $\text{SBP}_F$ and the latter half as $\text{SBP}_L$. In addition, the change $\text{SBP}_C$ is computed as follows:
\begin{equation}
\text{SBP}_C = |\text{SBP}_F - \text{SBP}_L|,
\end{equation}
where $|\cdot|$ obtains the absolute value. 
For blood glucose, we conduct data augmentation by computing the gap between the mean values of blood glucose measurements before and after meals, which is denoted as $\text{BG}_C$ (see Table \ref{tab:attributes}). 

\begin{table}[!t]
\caption{Popular activities conducted by participants}
\centering
\begin{tabular}{llr}
\hline
Activity        & \#Entries & \multicolumn{1}{l}{Intensity code \cite{taylor1978questionnaire} }\\ \hline
Housework       & 343     & 2.5                                                                                \\
Walking         & 315     & 3.5                                                                                \\
Jogging         & 235     & 6.0                                                                                \\
Aerobic workout & 130     & 8.0                                                                                \\
Cycling         & 94      & 4.0                                                                                \\
Swimming        & 77      & 6.0                                                                                \\
Elliptical      & 46      & \multirow{2}{*}{6.0}                                                               \\
Gym             & 29      &                                                                                    \\ \bottomrule
\end{tabular}
\label{activity_summary}
\end{table}


Physical activity and exercises may reflect one's physical conditions and the general health status. Knapik et al. \cite{knapik2019relationship} showed that people with chronic diseases generally have a lower level of physical fitness than healthy people. We list the most frequently uploaded activities in Table \ref{activity_summary}. As shown, the top three activities are housework, walking, and jogging. This is consistent with the profile of our participants that we have a large number of female elderly. In our study, instead of capturing the intensity of all activities, we group all the activities performed and uploaded by less than 50 participants during the entire 3-month study period as ``Others" (see Elliptical and Gym in Table \ref{activity_summary}). Then we compute the physical activeness of each individual participant, denoted as $A$, using the widely adopted intensity formula \cite{taylor1978questionnaire}:

\begin{equation}\label{equ:fusion}
A = \sum{I_i\cdot F_i\cdot T_i},
\end{equation}
where $i$ denotes the $i$th activity, $I_i$ denotes the intensity code of activity $i$ which represents the activity's level of energy expenditure, $F_i$ denotes the number of sessions (i.e., days) that the participant performed activity $i$ during the 3-month study, and $T_i$ denotes the average time spent in each session.

\subsubsection{Missing Value Imputation}
Like most datasets in the healthcare domain, our collected S3D dataset also has missing attribute values. Directly deleting data entries having missing values will significantly reduce the size of the dataset. To address this issue, various missing values imputation methods \cite{lin2020missing, donders2006gentle} have been proposed, among which Mean Imputation (MI) is a popular method. In addition, Luo et al. \cite{luo2022missing} compared the performance of different missing value imputation methods on the diabetes prediction task and their experimental results showed that the KNN Imputation (KNNI) method outperforms the others. Therefore, for all experiments of this work, we apply both MI (across all participants) and KNNI (only across $k$ nearest neighbors) methods. Among all the 35 data features, only six of them have no missing value, namely age, gender, race, BMI, smoking habit, and drinking habit  (see the last column of Table \ref{activity_summary}). Therefore, we use the first four: age, gender, race, and BMI, as the anchoring features for KNNI, and set $k$ to 200 based on the preliminary results.
The details of the two imputation methods are as follows,
\begin{itemize}
\item Mean Imputation (MI): This method fills missing values by inserting the mean value of all non-missing values for the corresponding attribute \cite{donders2006gentle}. 
\item k-Nearest Neighbor Imputation (KNNI): This method first measures the distance (dissimilarity) among data instances to search $k$ nearest neighbors for each instance with missing value(s). Then, missing values of each attribute are imputed by inserting the average value or the majority class of the identified $k$ nearest neighbors.
\end{itemize}

\subsubsection{One-hot Encoder}
Before training ML models, we need to properly convert the categorical variables into a numeric format.
In this work, we utilize the one-hot encoder to create a new variable for each categorical feature and map each category value into a vector of 0s and one 1, where 1 represents the corresponding category belongingness.


\begin{sidewaystable}[thp]
\centering
\caption{Performance comparison on 3H diagnosis}
\begin{tabular}{lllllllll}
\hline
Disease                         & ML model             & Imputation & Accuracy                        & F1                              & Recall                          & Precision                       & TPR                             & TNR                             \\ \hline
\multirow{8}{*}{Diabetes}       & \multirow{2}{*}{XGB} & KNN        & \textbf{0.802} & \textbf{0.795} & \textbf{0.802} & \textbf{0.794} & \textbf{0.683} & 0.898                           \\
                                &                      & Mean       & 0.753                           & 0.743                           & 0.753                           & 0.754                           & 0.561                           & 0.898                           \\\cline{2-9}
                                & \multirow{2}{*}{RF}  & KNN        & 0.762                           & 0.754                           & 0.762                           & 0.753                           & 0.634                           & 0.864                           \\
                                &                      & Mean       & 0.743                           & 0.730                           & 0.743                           & 0.735                           & 0.561                           & 0.881                           \\\cline{2-9}
                                & \multirow{2}{*}{SVM} & KNN        & 0.584                           & 0.431                           & 0.584                           & 0.341                           & 0.000                           & \textbf{1.000} \\
                                &                      & Mean       & 0.584                           & 0.431                           & 0.584                           & 0.341                           & 0.000                           & \textbf{1.000} \\\cline{2-9}
                                & \multirow{2}{*}{KNN} & KNN        & 0.594                           & 0.574                           & 0.594                           & 0.582                           & 0.342                           & 0.780                           \\
                                &                      & Mean       & 0.574                           & 0.551                           & 0.574                           & 0.551                           & 0.317                           & 0.763                           \\ \hline
\multirow{8}{*}{Hyperlipidemia} & \multirow{2}{*}{XGB} & KNN        & 0.673                           & 0.732                           & 0.804                           & 0.672                           & 0.804                           & 0.511                           \\
                                &                      & Mean       & 0.663                           & 0.717                           & 0.768                           & 0.672                           & 0.768                           & \textbf{0.533} \\\cline{2-9}
                                & \multirow{2}{*}{RF}  & KNN        & \textbf{0.713} & \textbf{0.782} & 0.929                           & \textbf{0.675} & 0.929                           & 0.444                           \\
                                &                      & Mean       & 0.663                           & 0.739                           & 0.857                           & 0.649                           & 0.857                           & 0.422                           \\\cline{2-9}
                                & \multirow{2}{*}{SVM} & KNN        & 0.555                           & 0.713                           & \textbf{1.000} & 0.555                           & \textbf{1.000} & 0.000                           \\
                                &                      & Mean       & 0.525                           & 0.688                           & 0.946                           & 0.541                           & 0.946                           & 0.000                           \\\cline{2-9}
                                & \multirow{2}{*}{KNN} & KNN        & 0.535                           & 0.624                           & 0.696                           & 0.565                           & 0.696                           & 0.333                           \\
                                &                      & Mean       & 0.515                           & 0.602                           & 0.661                           & 0.552                           & 0.661                           & 0.333                           \\ \hline
\multirow{11}{*}{Hypertension}  & \multirow{2}{*}{XGB} & KNN        & \textbf{0.812} & \textbf{0.829} & 0.868                           & \textbf{0.793} & 0.868                           & 0.750                           \\
                                &                      & Mean       & 0.713                           & 0.734                           & 0.755                           & 0.714                           & 0.755                           & 0.667                           \\\cline{2-9}
                                & \multirow{2}{*}{RF}  & KNN        & 0.772                           & 0.800                           & 0.868                           & 0.742                           & 0.868                           & 0.667                           \\
                                &                      & Mean       & 0.753                           & 0.783                           & 0.849                           & 0.726                           & 0.849                           & 0.646                           \\\cline{2-9}
                                & \multirow{2}{*}{SVM} & KNN        & 0.525                           & 0.688                           & \textbf{1.000} & 0.525                           & \textbf{1.000} & 0.000                           \\
                                &                      & Mean       & 0.525                           & 0.688                           & \textbf{1.000} & 0.525                           & \textbf{1.000} & 0.000                           \\\cline{2-9}
                                & \multirow{2}{*}{KNN} & KNN        & 0.515                           & 0.588                           & 0.660                           & 0.530                           & 0.660                           & 0.354                           \\
                                &                      & Mean       & 0.564                           & 0.651                           & 0.774                           & 0.562                           & 0.774                           & 0.333                           \\\cline{2-9} 
                                & Expert Rule (w/o MV)  & -          & 0.643                           & 0.704                           & 0.569                           & 0.923                           & 0.569                           & 0.861                           \\
                                & \multirow{2}{*}{Expert Rule  }         & Mean       & 0.670                           & 0.614                           & 0.460                           & 0.923                           & 0.460                           & \textbf{0.949} \\
                                  &  & KNN        & 0.670                           & 0.614                           & 0.460                           & 0.923                           & 0.460                           & \textbf{0.949} \\ \bottomrule
\multicolumn{9}{l}{\textit{Note}: The term ``w/o MV” means data entries with missing values are removed instead of imputed.}
\end{tabular}

\label{tab:main_result}

\end{sidewaystable}[thp]

\subsection{Machine Learning Algorithms Used for 3H Diagnosis} \label{sec_ML}
In this subsection, we introduce the following ML algorithms used for 3H diagnosis:
\subsubsection{Random Forest (RF)} RF \cite{breiman2001random} conducts prediction tasks by constructing a large number of decision trees and then selecting the class based on majority voting.

\subsubsection{XGBoost (XGB)} XGBoost \cite{chen2016xgboost} is a scalable ML model based on tree boosting. Similar to RF, XGBoost also achieves high accuracy by having a set of decision trees, which lowers their model interpretability.

\subsubsection{k-Nearest Neighbors (KNN)} KNN \cite{peterson2009k} is a distance-based supervised learning algorithm that classifies new data samples by observing the behavior of their $k$ nearest neighbors.

\subsubsection{Support Vector Machine (SVM)} SVM \cite{amari1999improving} aims to find the best hyperplane which makes the margin between different classes the largest.

\section{Experiments} \label{sec_experiment}
In this section, we introduce how we conduct experiments for 3H diagnoses and compare the results obtained using various imputation methods and ML models. Furthermore, we analyze the explainability of the model outputs with the aid of Shapley values.

\subsection{Experimental Settings}

For all experiments in this work, we adopt the standard $k$-fold cross-validation approach. Specifically, for each fold of cross-validation, the training dataset comprising $k$-1 folds of training samples is further divided into $k$ smaller sets. Then, a model is trained on $k$-1 sets and validated on the remaining set, and this process is repeated $k$ times that the $k$-1 folds of training samples have been used for validation once. Subsequently, the best hyperparameter values identified from the iterative validations are used to train the final model on all the $k$-1 folds of training data samples. Finally, we apply the trained final model to assess the corresponding fold of the testing data samples and report the averaged results of the $k$ folds. Besides, we apply stratified $k$-fold cross-validation to ensure approximately the same class ratios among all data splits. To evaluate and compare the performance of all ML models, we adopt six widely used metrics: Accuracy, F1-score (F1), Precision, Recall, True Positive Rate (TPR), and False Positive Rate (FPR).

\subsection{Expert Rule as Baseline} \label{subsec:er}
Although many chronic diseases are normally diagnosed by conducting clinical tests (e.g., blood test for diabetes and hyperlipidemia), few of them can be diagnosed in non-clinical settings. For instance, blood pressure (BP) can be measured easily at home, and used to self-diagnose hypertension according the following expert rule:

\emph{If at least two readings of the systolic or diastolic blood pressure on two different occasions are more than 140 mmHg or 90 mmHg, respectively, the person can be diagnosed with hypertension.}

Therefore, other than comparing the results obtained by the various ML models, we also compare them with the expert rule for diagnosing hypertension (see the last three rows of Table \ref{tab:main_result}). 


\subsection{Experimental Results} \label{exp_result}
Table~\ref{tab:main_result} reports the model performance of XGB, RF, SVM, and KNN for 3H diagnoses. We have the following key observations from Table~\ref{tab:main_result}:

\begin{figure}[!t]
\centering
\includegraphics[width=0.7\columnwidth]{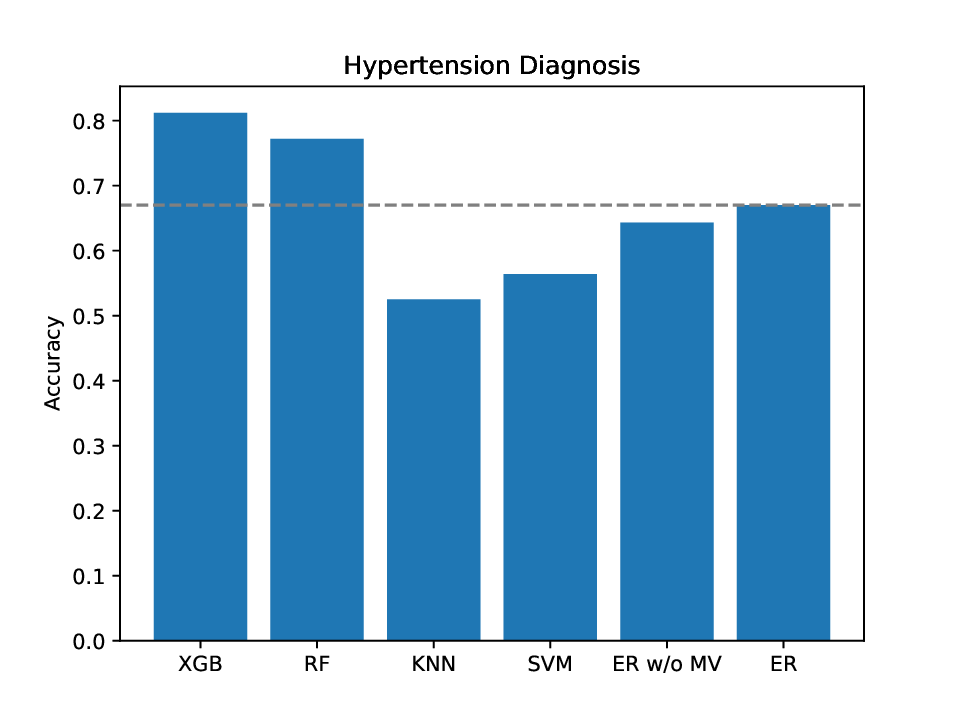}
\caption{Comparison of accuracy for hypertension diagnosis. ER denotes the expert rule (see Section \ref{subsec:er}), and MV denotes missing values.}
\label{figure:hyper_acc}
\end{figure}

(i) Overall speaking, XGB with KNNI achieves the best performance on diabetes and hypertension predictions. RF with KNNI outperforms the others on hyperlipidemia prediction. XGB and RF perform better than SVM and KNN in general. These findings are consistent with the literature and within expectation that ensemble methods generally perform better than single (shallow) learners in the healthcare domain \cite{luo2022missing}. 

(ii) As shown in Table~\ref{tab:main_result}, we also compare the best performance of each model with the expert rule on hypertension diagnosis. The results show that XGB and RF achieve better performance and outperform the expert rule (see Fig.~\ref{figure:hyper_acc}). Such finding is extremely encouraging because it demonstrates that certain disease diagnoses using comprehensive daily behavioral data as complements to the relevant measurement data may be more reliable than only using those measurement data.

(iii) Comparing between the two imputation methods, we can see that KNNI outperforms MI. This is consistent with the prior study on diabetes prediction using medical data \cite{luo2022missing} that for disease predictions, it is more accurate to impute the missing values only according to the most similar peers. 


\begin{figure*}
    \begin{subfigure}[b]{0.7\textwidth}
        \centering
        \includegraphics[width=\textwidth]{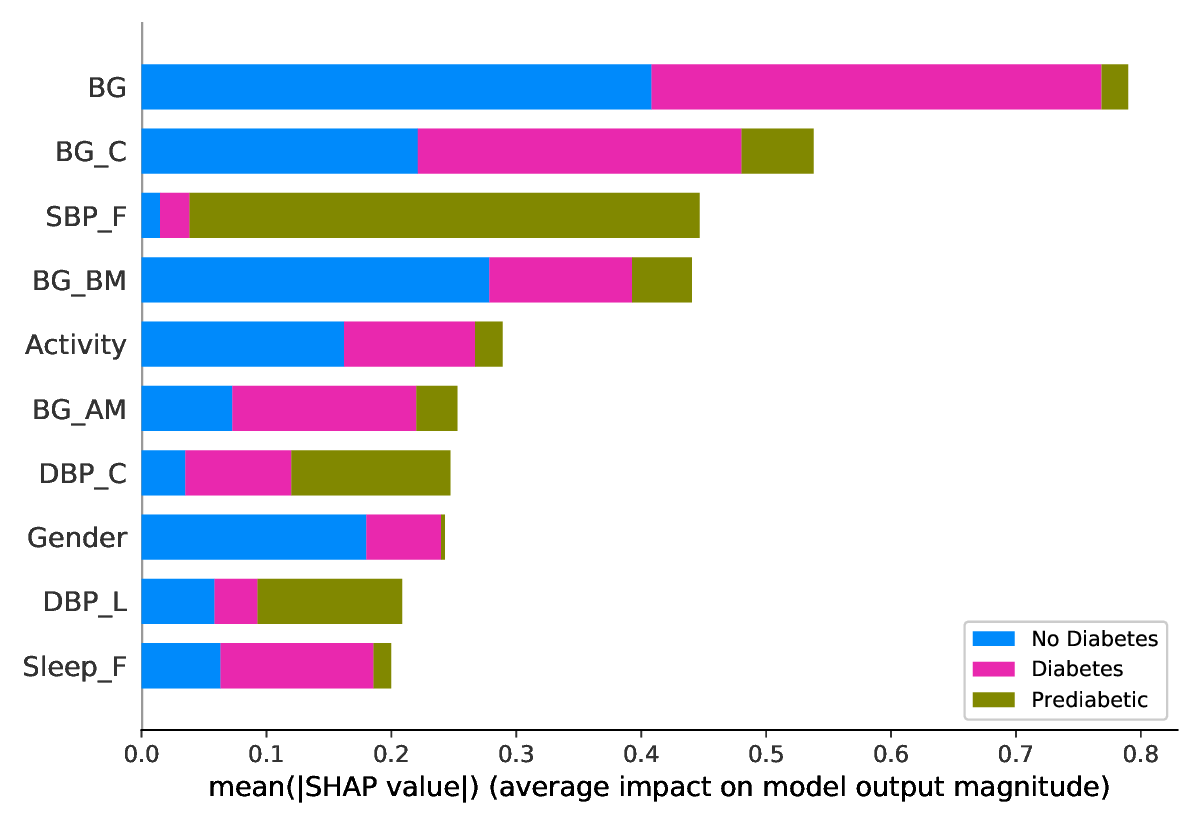}
        \caption{Diabetes}
    \end{subfigure}\hfill
    \begin{subfigure}[b]{0.7\textwidth}
        \centering
        \includegraphics[width=\textwidth]{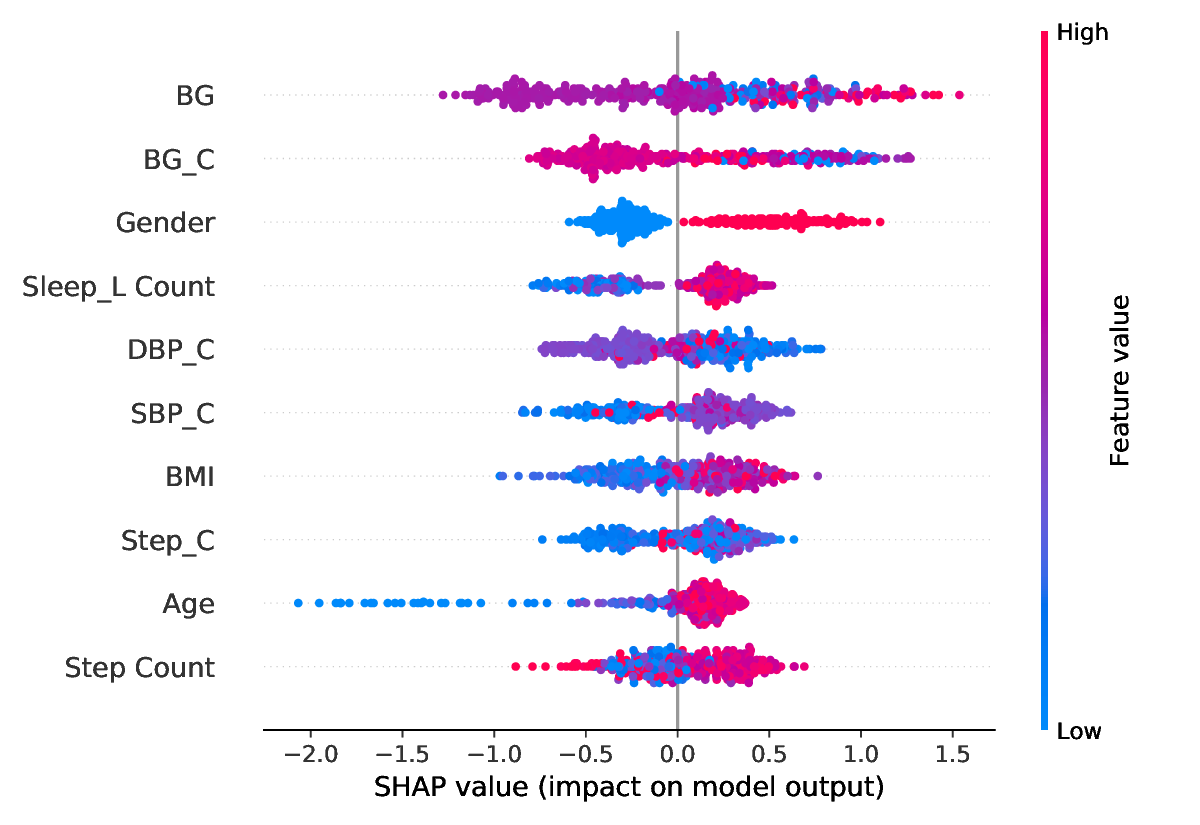}
        \caption{Hyperlipidemia}
        \centering
    \end{subfigure}\hfill
    \begin{subfigure}[b]{0.7\textwidth}
        \centering
        \includegraphics[width=\textwidth]{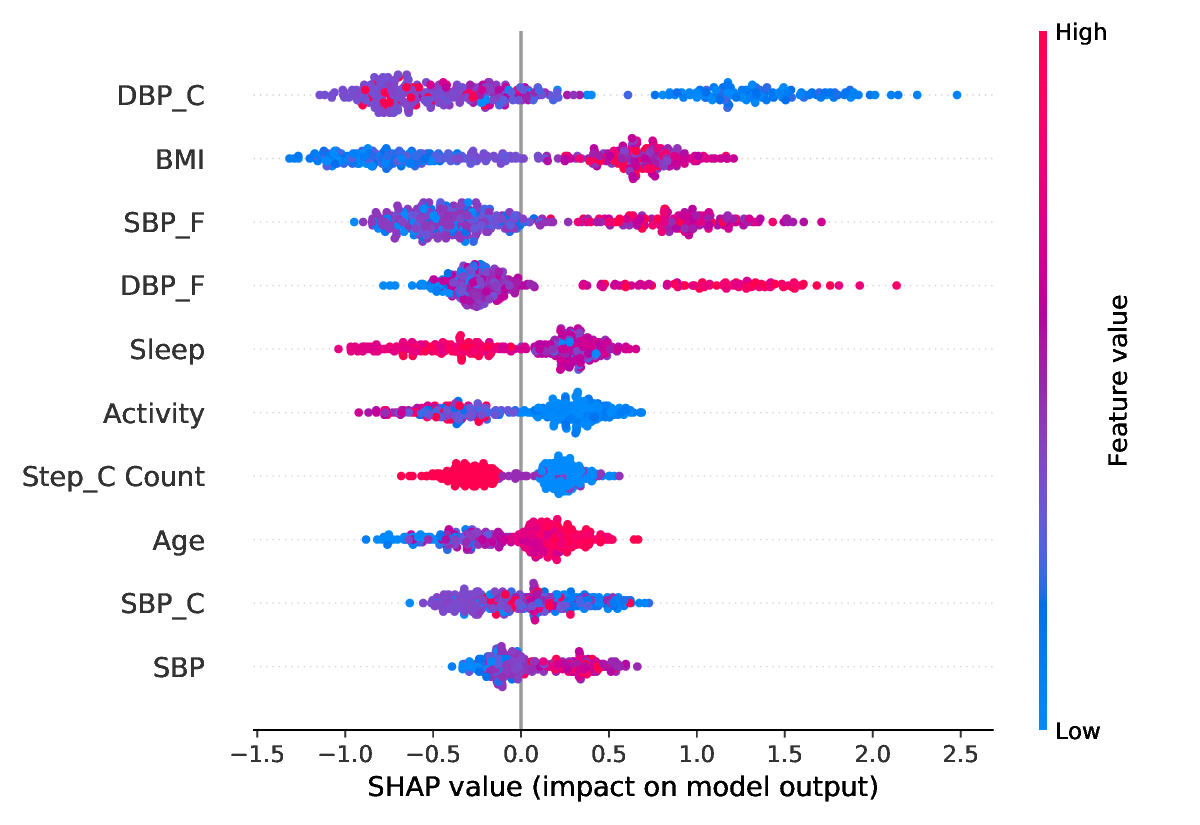}
        \caption{Hypertension}
        \centering
    \end{subfigure}
    
\caption{SHAP summary plot on hyperlipidemia and hypertension diagnoses using XGB.}
\label{fig:shap_hh}
\end{figure*}

\subsection{Shapley Analysis for Explainable 3H Diagnosis}\label{sec_exp_shap}
In this work, a total of 35 attributes are used for 3H diagnosis. To further analyze the importance of the individual attributes and provide explanations of the trained models, we leverage the Shapley values \cite{winter2002shapley} to investigate the contribution of each attribute to the diagnosis result.

Fig.~\ref{fig:shap_hh}(a) shows the SHAP summary plot for diabetes diagnosis using the overall best performing ML model XGBoost (XGB). The x-axis represents the magnitude of the Shapley values, and the top ten most influential attributes are sequentially listed along the y-axis. As shown in Fig.~\ref{fig:shap_hh}(a), the most influential attribute is the averaged self-reported blood glucose measurement. This finding is as expected that in clinical settings, glucose readings obtained from blood test results alone are used to diagnose diabetes \cite{assegie2022extraction}. Nonetheless, similar to hypertension diagnosis (see Section \ref{exp_result}), if we only use the series of ``BG" readings to diagnose diabetes, we end up with inferior results than those reported in Table~\ref{tab:main_result}.
The width of each colored horizontal bar roughly represents the attribute's contribution to the respective diabetes class. For instance, as shown in Fig.~\ref{fig:shap_hh}(a), ``BG'' plays a critical role in both diabetes-free and diabetes cases and has little influence on pre-diabetes \cite{assegie2022extraction}. Other than the ``BG" series, one's physical activeness (i.e., ``Activity") is the most influential attribute in diabetes prediction \cite{knapik2019relationship}. The top contributing attributes to distinguish pre-diabetes are blood pressure attributes and lifestyle (i.e., sleep hours). All these findings are consistent with the respective medical and commonsense knowledge, demonstrating the correctness of the trained XGBoost model.



Furthermore, we show SHAP summary plots for hyperlipidemia and hypertension diagnoses in Fig.~\ref{fig:shap_hh}. As shown in Fig.~\ref{fig:shap_hh}(b), comparing to blood pressure (BP) attributes, blood glucose (BG) measurements are more indicative attributes for hyperlipidemia diagnosis \cite{petrie2018diabetes}. Moreover, BMI and $\text{Step}_{C}$ have positive impact leading to hyperlipidemia, which means if a person can reduce his/her BMI and walk around more, he/she will highly likely have a lower risk of suffering from hyperlipidemia. Fig.~\ref{fig:shap_hh}(c) illustrates that the series of ``BP" attributes and BMI have the highest impact on hypertension diagnosis \cite{zhu2021prediction}. From the top three most influential ``BP" attributes, namely $\text{DBP}_{C}$, $\text{SBP}_{F}$, and $\text{DBP}_{F}$, we can see that the change in diastolic readings across the study period (i.e., $\text{DBP}_{C}$,) has a large amount of negative influence towards hypertension. In addition, if $\text{SBP}_{F}$ or $\text{DBP}_{F}$ is high, one may have a high possibility of suffering from hypertension, which is nothing more than commonsense. Similar to hyperlipidemia, one may have a higher chance to stay away from hypertension if he/she can reduce his/her BMI and be more active \cite{he2020prevalence}.


In summary, the Shapley analysis reveals the key influential attributes of the 3H diseases and provides explanations of why an ML model produces that output. These findings are highly consistent with both medical and commonsense knowledge, demonstrating the correctness of our approach in using comprehensive behavioral data (other than the accuracy shown in Table~\ref{tab:main_result}).

\section{Conclusion}
In this paper, we introduce how we diagnose 3H diseases using own collected behavioral data instead of using medical data collected in clinical settings. The accurate diagnosis results are encouraging which suggest that we may be able to detect the 3H diseases in their early stage by merely leveraging the daily behavioral data. 

Our own collected S3D dataset also comprises dietary information other than those listed in Table~\ref{tab:attributes} that many participants uploaded their three meals and snacks taken each day. Nonetheless, to complement the behavioral data with dietary habit, much effort is required on food ingredient decomposition and nutrition analysis. We leave this part as future work to further elevate 3H diagnosis accuracy.

\section{Acknowledgement}
This research is supported, in part, by the National Research Foundation (NRF), Singapore under its AI Singapore Programme (AISG Award No: AISG-GC-2019-003). Any opinions, findings and conclusions or recommendations expressed in this material are those of the authors and do not reflect the views of National Research Foundation, Singapore.

 \bibliographystyle{elsarticle-harv} 
 \bibliography{ref}





\end{document}